\documentclass[preprint,amsmath,amssymb,superscriptaddress, floatfix,nofootinbib]{revtex4}
\usepackage{epsfig,subfigure}
\usepackage{epstopdf}
\usepackage[utf8]{inputenc}
\usepackage{graphicx}
\usepackage{amssymb}
\usepackage{amsxtra}
\usepackage{amsmath}
\usepackage{booktabs,multirow,tabularx}
\usepackage{slashed}
\usepackage{float}
\usepackage{placeins}
\usepackage{rotating}
\usepackage{lscape}
\usepackage{color}
\usepackage{hyperref}
\usepackage{soul}

\newcommand{\bea}{\begin{eqnarray}}
\newcommand{\eea}{\end{eqnarray}}

\def\beq{\begin{equation}}
\def\eeq{\end{equation}}

\DeclareRobustCommand*{\ora}{\overrightarrow}

\DeclareUnicodeCharacter{2212}{-}

\begin{document}

\title{Higgs Production in Association with a Dark-$Z$ at\\ Future Electron Positron Colliders}

\author{Pierce Giffin}
\email{pgiffin@ucsc.edu}
\affiliation{Department of Physics and Astronomy, University of Kansas, Lawrence, Kansas 66045~ U.S.A.}
\affiliation{Department of Physics, University of California, Santa Cruz, CA 95064, USA}

\author{Ian M. Lewis}
\email{ian.lewis@ku.edu}
\affiliation{Department of Physics and Astronomy, University of Kansas, Lawrence, Kansas 66045~ U.S.A.}

\author{Ya-Juan Zheng}
\email{yjzheng@ku.edu}
\affiliation{Department of Physics and Astronomy, University of Kansas, Lawrence, Kansas 66045~ U.S.A.}


\begin{abstract}
In recent years there have been many proposals for new electron-positron colliders, such as the Circular Electron-Positron Collider, the International Linear Collider, and the Future Circular Collider in electron-positron mode.  Much of the motivation for these colliders is precision measurements of the Higgs boson and searches for new electroweak states.  Hence, many of these studies are focused on energies above the $h\,Z$ threshold.  However, there are proposals to run these colliders at the lower $WW$ threshold and $Z$-pole energies.  In this paper, we propose a new search for Higgs physics accessible at lower energies: $e^+e^-\rightarrow h\,Z_d$, where $Z_d$ is a new light gauge boson such as a dark photon or dark-$Z$.  Such searches can be conducted at the $WW$ threshold, i.e. energies below the $h\,Z$ threshold where exotic Higgs decays can be searched for in earnest.  Additionally, due to very good angular and energy resolution at future electron-positron colliders, these searches will be sensitive to $Z_d$ masses below 1 GeV, which is lower than the current direct LHC searches.  We will show that at $\sqrt{s}=160$ GeV with 10 ab$^{-1}$, a search for $e^+e^-\rightarrow h\,Z_d$ is sensitive to $h-Z-Z_d$ couplings of $\delta\sim 8\times 10^{-3}$ and cross sections of $\sim 1-2$~ab for $Z_d$ masses below 1 GeV.  The results are similar at $\sqrt{s}=240$ GeV with 5 ab$^{-1}$.

\end{abstract}

\maketitle

\section{Introduction}
\label{sec:intro}
A central aspect to many of the proposals for future high energy colliders after the LHC is the prospect of discovering new physics associated with the Higgs boson~\cite{Arkani-Hamed:2015vfh,Mangano:2016jyj,Cepeda:2019klc,deBlas:2019rxi,Strategy:2019vxc}.  The high luminosity and clean environments of future electron-positron colliders make them particularly well suited to searches for new physics in precision measurements of the SM Higgs~\cite{Bechtle:2014ewa,Liu:2016zki,Gu:2017ckc,Durieux:2017rsg,Barklow:2017suo,Voigt:2017vfz,An:2018dwb,deBlas:2018mhx,deBlas:2019wgy,Tan:2020fxk,Jung:2020uzh,Fuchs:2020cmm,Li:2020glc} and new electroweak (EW) physics in general.  Most of these studies of Higgs searches and measurements have been focused on energies at or above 240-250 GeV, which are the energies appropriate for EW production of the 125 GeV Higgs boson in the SM.  Indeed, there has been much work on discovering new states associated with the Higgs boson at these energies~\cite{Drechsel:2018mgd,Kalinowski:2018kdn,Bahl:2020kwe}.  

However, many proposed lepton colliders such as the Circular Electron Positron Collider (CEPC)~\cite{CEPCStudyGroup:2018rmc,GuimaraesdaCosta:2678417}, International Linear Collider (ILC)~\cite{Behnke:2013xla,Baer:2013cma,Adolphsen:2013jya}, and Future Circular Collider in electron-positron mode (FCC-ee)~\cite{Abada:2019zxq} also propose to conduct precision measurements of EW parameters with runs at the $Z$-pole and/or $WW$-threshold.  In this paper, we propose a novel search involving the 125 GeV Higgs boson that can be carried out at energies below 240 GeV: Higgs production in association with a new light gauge boson. 

Many models have the Higgs boson ($h$) as a portal to a dark sector with couplings to a gauge boson of a broken $U(1)$ symmetry.  Such gauge bosons are theoretically well-motivated (see Ref.~\cite{Langacker:2008yv,Jaeckel:2010ni,Hewett:2012ns,Essig:2013lka,Alexander:2016aln,Battaglieri:2017aum} and references therein) and are the so-called ``dark photons''~\cite{Holdom:1985ag,Galison:1983pa,Dienes:1996zr,Bjorken:2009mm} or ``dark-$Z$s''~\cite{Davoudiasl:2012qa,Davoudiasl:2013aya,Davoudiasl:2012ag}.  If the new gauge boson ($Z_d$) is light enough the Higgs boson can decay into it: $h\rightarrow Z\,Z_d$~\cite{Davoudiasl:2012ag,Davoudiasl:2013aya,Davoudiasl:2015bua,Curtin:2013fra,Curtin:2014cca} and $h\rightarrow Z_dZ_d$~\cite{Gopalakrishna:2008dv,Curtin:2013fra,Curtin:2014cca}.  The same interaction that facilitates the $h\rightarrow Z\,Z_d$ decay can also facilitate Higgs production in association with the $Z_d$: $f\bar{f}\rightarrow Z^*\rightarrow h\,Z_d$.  In this paper we will show that $e^+e^-\rightarrow Z^*\rightarrow h\,Z_d$ is possibly discoverable at future electron-positron colliders.    There are two interesting points to make about this search:
\begin{enumerate}
\item If the $Z_d$ is light, $e^+e^-\rightarrow h\,Z_d$ can be searched for at colliders with energies below 240 GeV.  Indeed, we will show that this process is discoverable at a 160 GeV machine with 10 ab$^{-1}$ of data~\cite{Abada:2019zxq}, i.e. the $WW$ threshold.  Hence, it is possible to search for exotic $h-Z-Z_d$ couplings below energies for which the Higgs can be produced via the traditional mode $e^+e^-\rightarrow h\,Z$, at which point exotic Higgs decays can be searched for in earnest.
\item As we also will show, due to the very good energy and angular resolutions, for leptonic $Z_d\rightarrow \ell^+\ell^-$ decays it will be possible to discover a dark-$Z$ with mass below 1 GeV, $m_{Z_d}\lesssim 1$~GeV, via its interactions with the Higgs.  Current LHC searches for $h\rightarrow Z\,Z_d$ and $h\rightarrow Z_d\,Z_d$ are limited to masses above 1 GeV~\cite{Aaboud:2018fvk,CMS-PAS-HIG-19-007}.  Hence, this search at a future electron-positron collider has the potential to open up new regions of parameter space.
\end{enumerate}

The paper is organized as follows. In Sec.~\ref{sec:Model} we introduce the model and give an overview of current constraints on model parameters.  We present our collider analysis in Sec.~\ref{sec:collider}.  The results of our study are given in Sec.~\ref{sec:results}.  Finally, in Sec.~\ref{sec:conc} we conclude.

\section{Model and Constraints}
\label{sec:Model}
We consider a massive vector boson, $Z_d$, originating from a broken $U(1)_d$ and assume the SM fermions are not charged under this $U(1)_d$.  Hence, the Higgs will serve as the portal between a dark sector and the SM.  We will consider interactions after $SU(2)_L\times U(1)_Y$ breaking.  There are two scenarios of interest for us, when the $Z_d$ is a  dark photon~\cite{Holdom:1985ag,Galison:1983pa,Dienes:1996zr,Bjorken:2009mm} or dark-$Z$~\cite{Davoudiasl:2012qa,Davoudiasl:2013aya,Davoudiasl:2012ag}.  In the dark photon scenario, the $Z_d$ kinetically mixes with the SM hypercharge.  This scenario has garnered much attention in the literature and there have been many searches at low energy, high intensity experiments~\cite{Alexander:2016aln,Battaglieri:2017aum}.  The kinetic mixing is often thought of as arising via loops of particles charged under both the SM hypercharge and the $U(1)_d$ symmetry.  If the particles inside the loops couple to the Higgs boson, these loops can also induce effective interactions between the Higgs and dark-photon~\cite{Davoudiasl:2012ig}:\footnote{We follow the notation of Ref.~\cite{Davoudiasl:2013aya} for the operators.}
\begin{eqnarray}
\mathcal{O}_{B,X}=\frac{c_{B,X}}{\Lambda}\,h\,X_{\mu\nu} Z^{\mu\nu}_d,\quad{\rm where}\quad X=Z,\gamma,Z_d,\label{eq:CB}
\end{eqnarray}
and $\Lambda$ is a new physics scale.
An alternative is the dark-$Z$ scenario~\cite{Davoudiasl:2012qa,Davoudiasl:2013aya,Davoudiasl:2012ag} when the SM $Z$ and the $Z_d$ have mass mixing.  This case typically has a different coupling pattern to fermions than the dark photon, and can indeed induce new sources of parity violation~\cite{Davoudiasl:2012qa,Davoudiasl:2012ag}.  In this case the couplings with the Higgs are expected to occur at dimension 3:
\begin{eqnarray}
\mathcal{O}_{A,X}=c_{A,X}h\,X_\mu\,Z^\mu_d,\quad{\rm where}\quad X=Z,Z_d.\label{eq:CA}
\end{eqnarray}
Here we cannot have $X=\gamma$ due to gauge invariance.  Interactions such as $\mathcal{O}_{A,X}$ can also be induced via the mixing of the SM Higgs with the Higgs of a dark sector, see Ref.~\cite{Gopalakrishna:2008dv} for an example.  

In this paper we propose to search for $e^+e^-\rightarrow X^*\rightarrow h\,Z_d$.  Unless the SM is charged under the $U(1)_d$, the coupling $e^+-e^--Z_d$ will be suppressed by a mixing angle.  Hence, this rate is suppressed for $X=Z_d$.  If $X=Z$, the rate is not suppressed by the initial state coupling.  The photon mediated process $X=\gamma$ is also possible for the operator $\mathcal{O}_{B,X}$, but not possible for $\mathcal{O}_{A,X}$.  

\begin{figure}[tb]
\begin{center}
\includegraphics[width=0.5\textwidth,clip]{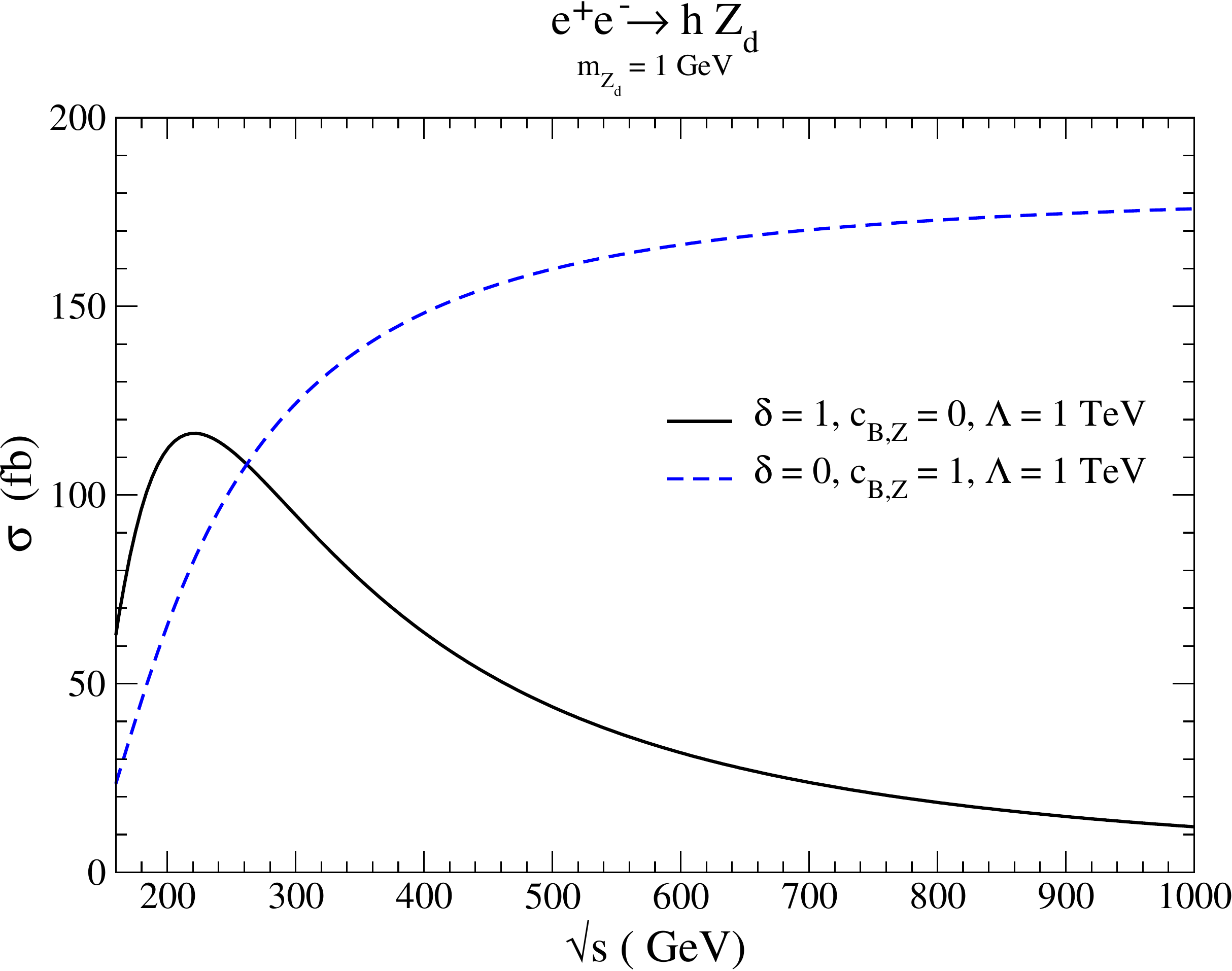}
\end{center}
\caption{\label{fig:Prod} Production cross section for $e^+e^-\rightarrow h\,Z_d$ as a function of center of momentum energy $\sqrt{s}$ with (black solid) $\delta=1,\,c_{B,Z}=0$ and (blue dashed) $\delta=0,\,c_{B,Z}=1$.  For both curves $m_{Z_d}=1$~GeV and $\Lambda=1$~TeV.}
\end{figure}

From this discussion, we will consider interactions with $X=Z$.  Following Refs.~\cite{Davoudiasl:2012ag,Davoudiasl:2013aya}, we parameterize $c_{A,Z}$ assuming that $\mathcal{O}_{A,Z}$ arises from $Z-Z_d$ mass mixing: 
\begin{eqnarray}
c_{A,Z}=\frac{g}{c_W}m_{Z_d}\,\delta.\label{eq:CAZ}
\end{eqnarray}
We will also consider the limit where the mass of the $Z_d$ is much smaller than other scales in the problem: $m_{Z_d}\ll m_Z,\sqrt{s}$, where $m_Z$ is the $Z$ mass and $\sqrt{s}$ is center of momentum frame energy.  The production rate when only $\mathcal{O}_{A,Z}$ contributes is 
\begin{eqnarray}
\sigma_A(s)=\frac{g^4\,\delta^2}{384\,\pi\,c_W^4}\left(g_L^2+g_R^2\right)\frac{1}{s}\left(1-\frac{m_h^2}{s}\right)^3\left(1-\frac{m_Z^2}{s}\right)^{-2}+\mathcal{O}\left(\frac{m_{Z_d}^2}{s^2}\right)\label{eq:rateCA}
\end{eqnarray}
whereas the rate when only $\mathcal{O}_{B,Z}$ contributes is
\begin{eqnarray}
\sigma_B(s)&=&\frac{g^2\,c_{B,Z}^2}{48\,\pi\,c_W^2\,\Lambda^2}\left(g_L^2+g_R^2\right)\left(1-\frac{m_h^2}{s}\right)^3\left(1-\frac{m_Z^2}{s}\right)^{-2}+\mathcal{O}\left(\frac{m_{Z_d}^2}{s\,\Lambda^2}\right),\label{eq:rateCB}
\end{eqnarray}
where $m_h=125$~GeV is the Higgs boson mass, $g_L=s_W^2$ and the $g_R=-1/2+s_W^2$ are the left- and right-chiral electron-$Z$ couplings, and $c_W=\cos\theta_W,\,s_W=\sin\theta_W$ with $\theta_W$ being the weak mixing angle. As can be seen the energy dependence of $\sigma_A(s)$ and $\sigma_B(s)$ is different, due to the fact that $\mathcal{O}_{A,Z}$ is a dimension-3 operator while $\mathcal{O}_{B,Z}$ is a non-renormalizable dimension-5 operator.  In Fig.~\ref{fig:Prod} we show the $e^+e^-\rightarrow h\,Z_d$ cross section for both $\delta=0$ and $c_{B,Z}=0$.  The dimension-5 cross section asymptotes to a constant value at high energy, while the cross section from $\mathcal{O}_{A,Z}$ peaks at $\sqrt{s}\sim 220$~GeV and then decreases as energy increases.

  An interesting feature of this is that ratios of cross sections at different energies can help shed light on the origin of the $h-Z-Z_d$ interaction.  Considering $\sqrt{s}=160$ and 240~GeV, the ratio of rates for $c_{B,Z}=0$ and $\delta=0$ are
\begin{eqnarray}
\frac{\sigma_A(s=(160~{\rm GeV})^2)}{\sigma_A(s=(240~{\rm GeV})^2)}=0.55\quad{\rm and}\quad\frac{\sigma_B(s=(160~{\rm GeV})^2)}{\sigma_B(s=(240~{\rm GeV})^2)}=0.25,
\end{eqnarray}
respectively.  As we go out to ever higher energies this effect becomes more pronounced.

As mentioned earlier, typical direct LHC searches for $h\rightarrow Z\,Z_d$ are for $m_{Z_d}\gtrsim 1$~GeV~\cite{Aaboud:2018fvk,CMS-PAS-HIG-19-007}.  Hence, their constraints are not relevant for our parameter space of $m_{Z_d}\lesssim 1$~GeV.  However, for these light masses there are still many relevant lower energy searches and precision measurements~\cite{Battaglieri:2017aum}.  The bounds from these searches are usually reported in the dark-photon parameter space.  For $m_{Z_d}$ in the $0.1-1$~GeV mass range, the dark photon mixing parameter is limited to be be less than $\sim 10^{-3}$.  If the $Z_d$ is light enough it can contribute to exotic meson decays $K\rightarrow \pi Z_d$~\cite{Davoudiasl:2014kua} and $B\rightarrow K Z_d$~\cite{Davoudiasl:2012ag}, both of which place limits $|\delta| \lesssim 10^{-3}$.   There are a few caveats to these bounds.  First, there can be cancellations between the kinetic and mass mixings that can alleviate some of these bounds~\cite{Davoudiasl:2014kua}.  Additionally, many of these constraints are on the SM fermion and gauge boson couplings to the $Z_d$.  The connection between the fermion-$Z_d$ couplings and $c_{A,Z},c_{B,Z}$ will depend precisely on the UV complete model and the origin of $\mathcal{O}_{A,Z}$ and $\mathcal{O}_{B,Z}$. Due to this, it is interesting to independently search for processes that arise out of the $h-Z-Z_d$ couplings, even if the parameter space appears to be constrained. 

There are additional limits from the LHC that bound the $h-Z-Z_d$ coupling.  Using measurements of the Higgs production and decay modes via SM particles, it is possible to place limits on the Higgs branching ratio to undetected final states, ${\rm BR}_{\rm BSM}$, with minimal assumptions~\cite{Heinemeyer:2013tqa}.  These bounds are relevant for $h\rightarrow Z\,Z_d$ independent of dedicated searches.  Assuming no other new physics decay modes or modifications to couplings, the branching ratio into $h\rightarrow Z Z_d$ is
\begin{eqnarray}
{\rm BR}(h\rightarrow Z\,Z_d)=\frac{\Gamma(h\rightarrow Z\,Z_d)}{\Gamma_{\rm SM}+\Gamma(h\rightarrow Z\,Z_d)},
\end{eqnarray}
where $\Gamma_{\rm SM}=4.088$~MeV is the total width into SM final states~\cite{deFlorian:2016spz}.  Assuming $m_{Z_d}\ll m_h$ the partial width with $c_{B,Z}=0$ is
\begin{eqnarray}
\Gamma(h\rightarrow Z\,Z_d)=\frac{g^2\,\delta^2}{64\,\pi\,c_W^2}\frac{m_h^3}{m_Z^2}\left(1-\frac{m_Z^2}{m_h^2}\right)^3
\end{eqnarray}
and if $\delta=0$ the partial width is
\begin{eqnarray}
\Gamma(h\rightarrow Z\,Z_d)=\frac{1}{8\pi}\left(\frac{c_{B,Z}}{\Lambda}\right)^2m_h^3\left(1-\frac{m_Z^2}{m_h^2}\right)^3.
\end{eqnarray}
At the HL-LHC the projected limit is $\rm{BR}_{\rm BSM}\leq 0.025$~\cite{Cepeda:2019klc}.  This translates to a bound $|\delta|\leq 0.04$ or $|c_{B,Z}/\Lambda|\leq 0.11~{\rm TeV}^{-1}$.  As we will show, searches for the direct production $e^+e^-\rightarrow h\,Z_d$ are more sensitive to the $h-Z-Z_d$ coupling.

\section{Collider Analysis}
\label{sec:collider}
We study the process $e^+e^-\to Z^*\to h\,Z_d$ at the energies $\sqrt{s}=160$~GeV and $240$~GeV.    For simplicity, in our collider analysis we will assume only $\mathcal{O}_{A,Z}$ contributes and set
\begin{eqnarray}
c_{B,Z}=0.
\end{eqnarray}
For better signal reconstruction,the leptonic decays of the $Z_d$ are considered: $Z_d\rightarrow \ell^+\ell^-$ with $\ell=e,\mu$.  To increase rates, we consider the hadronic decays of the Higgs: $h\rightarrow b\bar{b},gg,c\bar{c}$.  \texttt{Madgraph5\_aMC@NLO}~\cite{Alwall:2014hca} is used for all event generation with the interaction $\mathcal{O}_{A,Z}$ implemented via ~\texttt{FeynRules}~\cite{Christensen:2008py,Alloul:2013bka}.  The Higgs branching ratios are normalized to agree with the LHC Higgs Cross Section Working Group~\cite{deFlorian:2016spz}.  Production and decays are simulated at parton level, then detector effects are modeled by Gaussian smearing of the energies and momentum of final state particles.  Final state quarks and gluons are smeared according to the hadronic calorimeter energy resolution~\cite{GuimaraesdaCosta:2678417,Abada:2019zxq}:
\begin{eqnarray}
\frac{\Delta E}{E}=\frac{0.34}{\sqrt{E/{\rm GeV}}}.\label{eq:hadsmear}
\end{eqnarray}
The electron and muon energies are smeared according to the tracking system momentum resolution~\cite{GuimaraesdaCosta:2678417,Abada:2019zxq}:
\begin{eqnarray}
\Delta\left(\frac{1}{p_T}\right)=2\times 10^{-5}\,{\rm GeV}^{-1}\otimes \frac{10^{-3}}{p_T\sqrt{\sin\theta}}.
\end{eqnarray}

We adopt the following pre-selection acceptance cuts:
\begin{gather}
p_{T,j}>20~{\rm GeV},\quad p_{T,\ell}>0.5~{\rm GeV}\nonumber\\
|\cos\theta_{\ell,j}|<0.98,\quad \Delta R_{ij}>0.005,\quad p_{T,\ell\ell}>20~{\rm GeV},
\end{gather}
where $p_{T,j},p_{T,\ell}$ are jet and lepton transverse momentum, $\theta_{\ell,j}$ are the lepton and jet polar angles in the detector, $\Delta R_{ij}=\sqrt{(\Delta \phi_{ij})^2+(\Delta \eta_{ij})^2}$ with $\Delta\phi_{ij}$ ($\Delta\eta_{ij}$) the azimuthal angle (rapidity) difference between particles $i$ and $j$, and $p_{T,\ell\ell}$ is the di-lepton transverse momentum.  For all numerical results presented in this section, we consider the signal parameter points:
\begin{eqnarray}
\delta\times\sqrt{{\rm BR}(Z_d\rightarrow \ell^+\ell^-)}=1.5\times10^{-2}\quad{\rm with}\quad m_{Z_d}=0.5~{\rm and}~1~\rm{GeV},
\end{eqnarray}
where ${\rm BR}(Z_d\rightarrow \ell^+\ell^-)={\rm BR}(Z_d\rightarrow \mu^+\mu^-)+{\rm BR}(Z_d\rightarrow e^+e^-)$.

\begin{figure}[tb]
\includegraphics[width=1\textwidth]{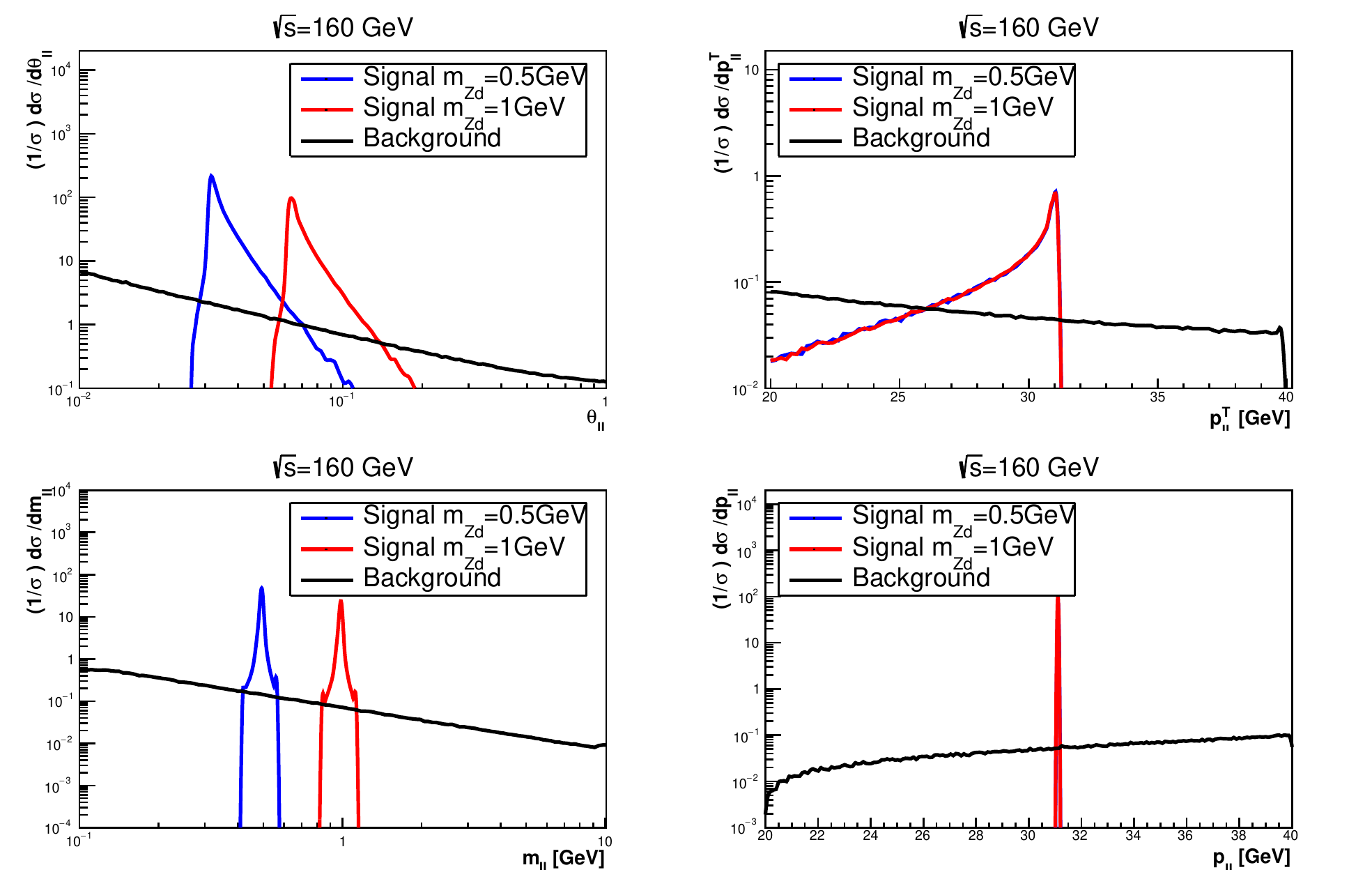}
\caption{\label{fig:160ll}Normalized distributions of the di-lepton (upper left) opening angle $\theta_{\ell\ell}$, (upper right) transverse momentum $p_{T,\ell\ell}$, (lower left) invariant mass $m_{\ell\ell}$, and (lower right) 3-momentum magnitude $|\ora{p}|_{\ell\ell}$ for (black) the $\ell^+\ell^-jj$ background, (red) the signal with $m_{Z_d}=1$~GeV, and (blue) the signal with $m_{Z_d}=0.5$~GeV.  These are at $\sqrt{s}=160$~GeV. 
}
\end{figure}

\begin{figure}[tb]
\includegraphics[width=1\textwidth]{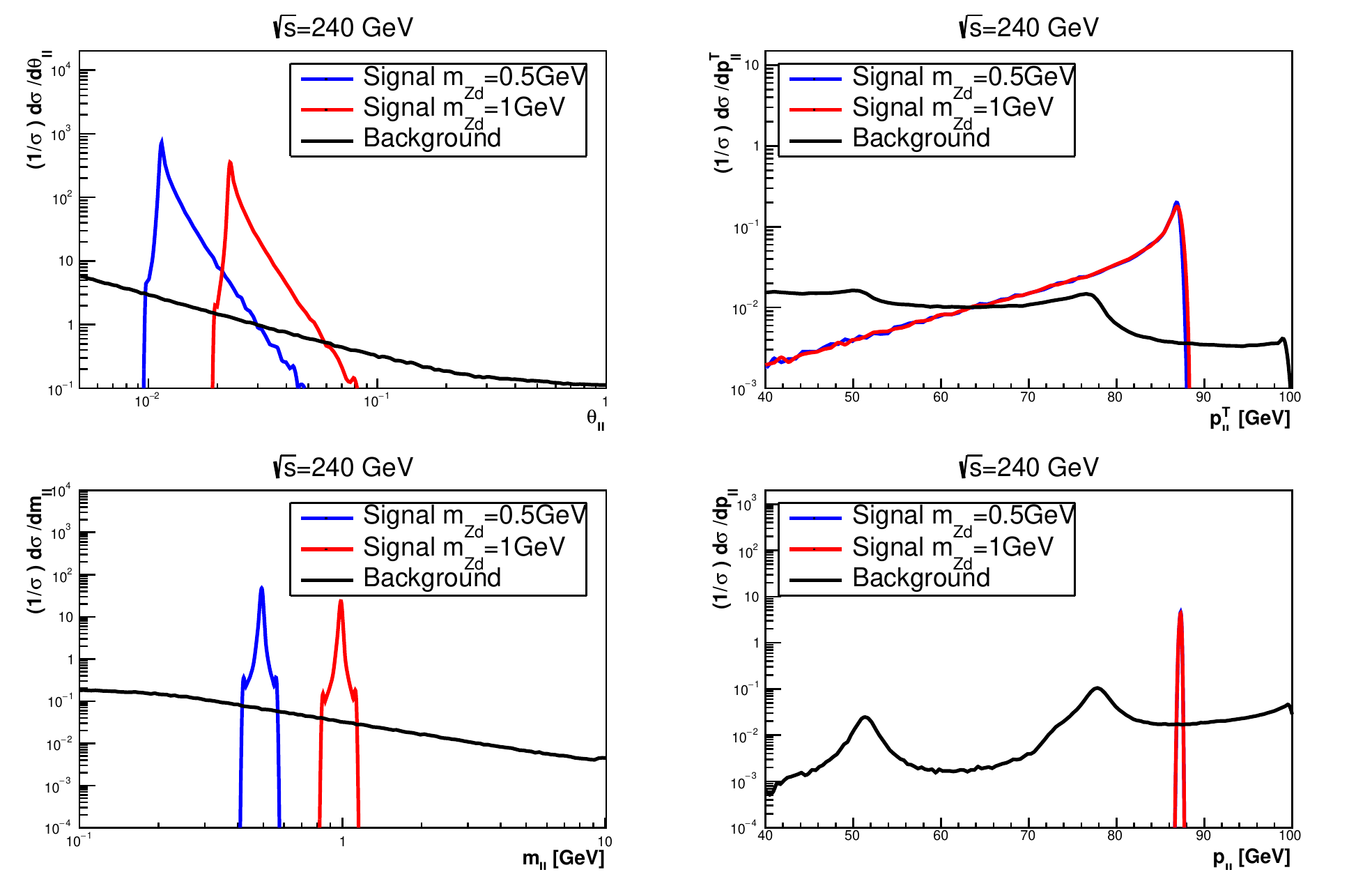}
\caption{\label{fig:240ll} Same as Fig.~\ref{fig:160ll} with $\sqrt{s}=240$~GeV.}
\end{figure}

For the background process, we include all the channels, except $h\,\ell^+\ell^-$, which can produce $\ell^+\ell^-jj$ and fully take into consideration the interference between the different SM background processes.  The background $h\,\ell^+\ell^-$ is considered separately at $\sqrt{s}=240$~GeV due to the very small Higgs width.  In Figs.~\ref{fig:160ll} and~\ref{fig:240ll} we show the normalized distributions of signal and background cross sections at $\sqrt{s}=160$ GeV and 240 GeV, respectively.  Clearly, the signal has very different behavior than the background.  The high energy and angular resolution of future lepton-colliders can help significantly to separate signal from background.  The angular resolution is expected to be better than $10^{-3}$ and the energy resolution for a lepton with energy below 100 GeV is $\lesssim 1\%$~\cite{GuimaraesdaCosta:2678417,Abada:2019zxq}.  Hence, from these figures it is clear that the decay products of the $Z_d$ are resolvable and the $Z_d$ mass is reconstructable even for $m_{Z_d}\lesssim 1$~GeV.

Signal versus background discrimination also benefits from the signal being a two-to-two process, while the background in the signal region is at best two-to-three.  From the $e^+e^-\rightarrow h\,Z_d$ kinematics, the energy of the $Z_d$ is expected to be
\begin{eqnarray}
E_{Z_d}=\frac{s+m_{Z_d}^2-m_h^2}{2\,\sqrt{s}}.
\end{eqnarray}
For $m_{Z_d}\lesssim 1$~GeV, the $Z_d$ mass can be neglected and the magnitude of the $Z_d$ three-momentum is 
\begin{eqnarray}
|\ora{p}_{Z_d}|=\begin{cases}31~{\rm GeV}\quad & {\rm for}~\sqrt{s}=160~{\rm GeV}\\
87~{\rm GeV}\quad & {\rm for}~\sqrt{s}=240~{\rm GeV}
\end{cases}.\label{eq:Zdmom}
\end{eqnarray}
This is precisely where the signal di-lepton three-momentum and transverse momentum peak in Figs.~\ref{fig:160ll} and~\ref{fig:240ll}.

\begin{table}[tb]
 \begin{center}
  \begin{tabular}{| c ||c||c|c||c|c|c|}
   \hline
\multicolumn{2}{|c||}{Cross Sections} &\multicolumn{2}{c||}{$\sqrt{s}=160$ GeV}&\multicolumn{3}{c|}{$\sqrt{s}=240$ GeV}\\\cline{1-2}\cline{3-4}\cline{5-7}
$m_{Z_d}$ &Cut&\multicolumn{1}{c|}{Signal}&\multicolumn{1}{c||}{Background}&\multicolumn{1}{c|}{Signal}&\multicolumn{2}{c|}{Background}  \\\cline{6-7}
   &Flow &&$\ell\ell jj$  &&$\ell\ell jj$   & $h\ell^+\ell^-$  \\
   \hline \hline
\multirow{6}{*}{0.5 GeV}& Pre-selection                     &8.8 ab&  7.6$\times 10^5$ ab    &16 ab&  7.4$\times 10^5$ ab  &1.0$\times 10^4$ ab \\ 
& $+m_{jj}$ cut                                             &8.2 ab&  2.1$\times 10^4$ ab    &15 ab&  1.2$\times 10^4$ ab  &  9.7$\times 10^3$ ab\\ 
& $+|\ora{p}|_{\ell\ell}$ cut                               &8.2 ab&  1.2$\times 10^3$   ab&15 ab&  2.6$\times 10^3$ ab   &5.2 ab \\ 
& $+p_{T,\ell\ell}$ cut                                     &6.4 ab&  290  ab &13 ab&  370  ab & 3.3 ab\\ 
& $+\theta_{\ell\ell}$ cut                                  &6.4 ab&  62 ab&13 ab&  100  ab & $7.9\times10^{-3}$ ab\\ 
& $+m_{\ell\ell}$ cut                                       &6.4 ab&  18 ab&13 ab&  55  ab  &$1.4\times10^{-3}$ ab\\ 
\hline\hline
\multirow{6}{*}{1 GeV}& Pre-selection                     &8.8 ab&  7.6$\times 10^5$ ab &16 ab&  7.4$\times 10^5$ ab & 1.0$\times 10^4$  ab\\ 
& $+m_{jj}$ cut                                           &8.2 ab&  2.1$\times 10^4$  ab&15  ab& 1.2$\times 10^4$  ab  & 9.7$\times 10^3$  ab  \\ 
& $+|\ora{p}|_{\ell\ell}$ cut                                     &8.2 ab&  1.2$\times 10^3$ ab  &15 ab&  2.6$\times 10^3$ ab    &5.2  ab\\ 
 & $+p_{T,\ell\ell}$ cut                                  &6.4 ab&  290 ab  &13 ab&  370 ab   & 3.3 ab\\ 
& $+\theta_{\ell\ell}$ cut                                &6.4 ab&69 ab&13 ab&100  ab & $4.1\times10^{-2}$  ab\\ 
& $+m_{\ell\ell}$ cut                                     &6.4 ab&21  ab&13 ab&28  ab  &$2.6\times10^{-3}$  ab\\ 
   \hline
  \end{tabular}
  \caption{ Cut flow table for both $\sqrt{s}=160$ and $240$~GeV for our benchmark point  $\delta\times\sqrt{{\rm BR}(Z_d\rightarrow \ell\ell)}=1.5\times 10^{-2}$.  Here ${\rm BR}( Z_d \rightarrow \ell^+\ell^-)={\rm BR}(Z_d\rightarrow e^-e^+)+{\rm BR}(Z_d\rightarrow \mu^-\mu^+)$. }\label{tab:cutflow}
 \end{center}
\end{table}

Based on these considerations, we adopt the following cuts.  At $\sqrt{s}=160$~GeV, we require
\begin{eqnarray}
30.5~{\rm GeV}<&|\ora{p}|_{\ell\ell}&<31.5~{\rm GeV},\\
25~{\rm GeV}<&p_{T,\ell\ell}&<31.5~{\rm GeV},\nonumber\\
0.051<\theta_{\ell\ell}<0.2~{\rm for}~ m_{Z_d}=1~ {\rm GeV},&&\quad 0.03<\theta_{\ell\ell}<0.1~{\rm for}~ m_{Z_d}=0.5~ {\rm GeV}, \nonumber\\
0.8<m_{\ell\ell}/{\rm GeV}<1.25~{\rm for}~ m_{Z_d}=1~ {\rm GeV},&&\quad 0.4<m_{\ell\ell}/{\rm GeV}<0.6~{\rm for}~ m_{Z_d}=0.5 ~{\rm GeV}.\nonumber
\end{eqnarray}
At $\sqrt{s}=240$~GeV, we require
\begin{eqnarray}
85~{\rm GeV}<&|\ora{p}|_{\ell\ell}&<90~{\rm GeV},\\
60~{\rm GeV}<&p_{T,\ell\ell}&<90~{\rm GeV},\nonumber\\
0.018<\theta_{\ell\ell}<0.089~{\rm for}~ m_{Z_d}=1~ {\rm GeV},&&\quad 0.009<\theta_{\ell\ell}<0.045~{\rm for}~ m_{Z_d}=0.5~ {\rm GeV}, \nonumber\\
0.75<m_{\ell\ell}/{\rm GeV}<1.25~{\rm for}~ m_{Z_d}=1~ {\rm GeV},&&\quad 0.25<m_{\ell\ell}/{\rm GeV}<0.75~{\rm for}~ m_{Z_d}=0.5~ {\rm GeV}.\nonumber
\end{eqnarray}
Since the signal also contains the hadronic decays of a Higgs boson, we require the di-jet invariant mass to pass the cut\footnote{It may be possible to reconstruct the Higgs recoil mass from the reconstructed $Z_d$ and place a tighter cut.}
\begin{eqnarray}
115~{\rm GeV}<&m_{jj}&<135~{\rm GeV},
\end{eqnarray}
at both 160 and 240 GeV.

In Table~\ref{tab:cutflow} we show our cut flows for our benchmark signal points and background at both $\sqrt{s}=160$ and $240$~GeV. Here we include the $h\,\ell^+\ell^-$ backgrounds at 240 GeV.  There are also $\tau\tau jj$ backgrounds with leptonic $\tau$ decays, but they are sub-leading due to the small $\tau$ leptonic branching ratios and the fact that these backgrounds could be further suppressed with missing energy cuts. As can be seen, our cuts have small effect on our signal acceptance, with an efficiency of $72-82$\%, but are very efficient at suppressing background.  Cuts on the di-jet invariant mass suppress non-Higgs backgrounds by over an order of magnitude.  The di-lepton 3-momentum and transverse momentum cuts suppress the background by nearly another order of magnitude.  The di-lepton opening angle and invariant mass cuts suppress the background by yet another order of magnitude.  Hence, even though background starts out at more than four orders of magnitude larger than signal, after cuts the signal to background ratio is a manageable $S/B\sim 0.2-0.5$.

\section{Results}
\label{sec:results}

\begin{table}[tb]
\begin{center}
\begin{tabular}{|c|c||c||ccc||ccc|}\hline
$\sqrt{s}$ & Luminosity &$m_{Z_d}$& \multicolumn{3}{c||}{no b-tag}& \multicolumn{3}{c|}{1 b-tag}\\\cline{4-9}
 &  &  & Signal & Background & $\sigma_{\rm disc}$ & Signal & Background & $\sigma_{\rm disc}$ \\\hline
\multirow{2}{*}{160 GeV} & \multirow{2}{*}{10 ab$^{-1}$} & 0.5 GeV & 64 & 180& 4.6& 43 & 23 &7.3 \\
& & 1 GeV& 64 & 210 & 4.2 & 43 & 28 & 6.8\\\hline\hline
\multirow{2}{*}{240 GeV} & \multirow{2}{*}{5 ab$^{-1}$} & 0.5 GeV & 67 & 280& 3.9 & 46 & 43 &6.1 \\
& & 1 GeV& 67 & 140 & 5.3 & 46 & 22 & 7.7\\\hline
\end{tabular}
\caption{\label{tab:sig} Number of signal and background events, and discovery significance [Eq.~(\ref{eq:disc})] at both $\sqrt{s}=160$ and $240$ GeV with benchmark luminosities $10$ and $5$ ab$^{-1}$, respectively.  Results shown for benchmarks $\delta\times\sqrt{{\rm BR}(Z_d\rightarrow \ell^+\ell^-)}=1.5\times 10^{-2}$ and $m_{Z_d}=0.5$ and $1$ GeV. Here ${\rm BR}(Z_d\rightarrow \ell^+\ell^-)={\rm BR}(Z_d\rightarrow e^-e^+)+{\rm BR}(Z_d\rightarrow \mu^-\mu^+)$.}
\end{center}
\end{table}

We now present the results of our analysis, and estimate the reaches of future electron-positron colliders.   In Table~\ref{tab:sig}, we show the estimated number of signal and background events as well as the discovery significance at 160 GeV with 10 ab$^{-1}$ and 240 GeV with 5 ab$^{-1}$, which are the design luminosities of the FCC-ee~\cite{Abada:2019zxq}. Using a Poisson likelihood
\begin{eqnarray}
L(x|n)=\frac{x^n}{n!}e^{-x},
\end{eqnarray}
for a given number of signal events $S$ and background events $B$, the significance for discovery is given by the likelihood ratio~\cite{Cowan:2010js}
\begin{eqnarray}
\sigma_{\rm disc}=\sqrt{-2\ln\left(\frac{L(B|B+S)}{L(S+B|S+B)}\right)}=\sqrt{2\left(\left(B+S\right)\log\left(1+\frac{S}{B}\right)-S\right)}\label{eq:disc}.
\end{eqnarray}
That is, discovery is defined as when the background only hypothesis is disfavored at 5-sigma, $\sigma_{\rm disc}\geq 5$, relative to the background plus signal hypothesis.

\begin{figure}[htb]
\begin{center}
\subfigure[]{\includegraphics[width=0.45\textwidth,clip]{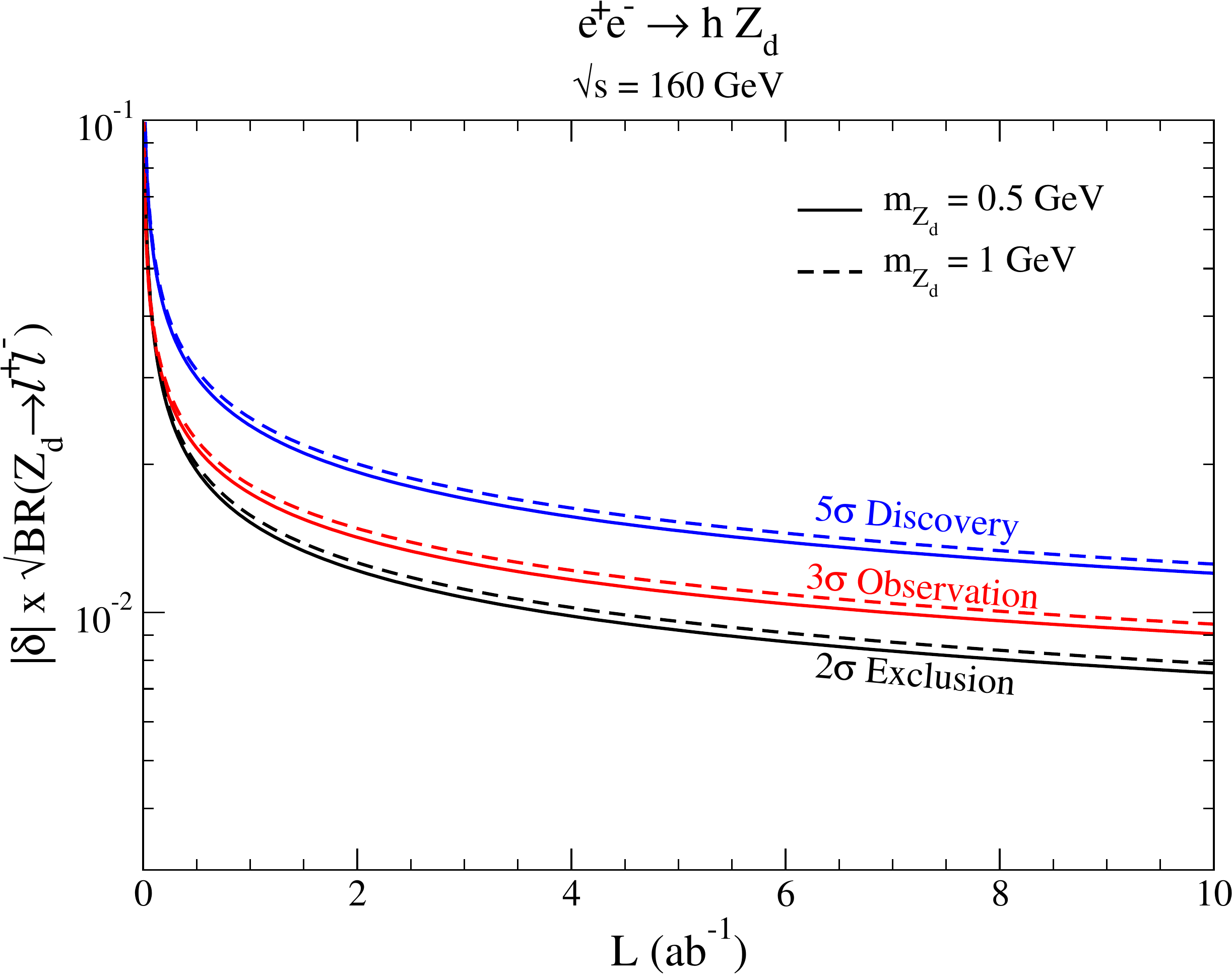}}
\subfigure[]{\includegraphics[width=0.45\textwidth,clip]{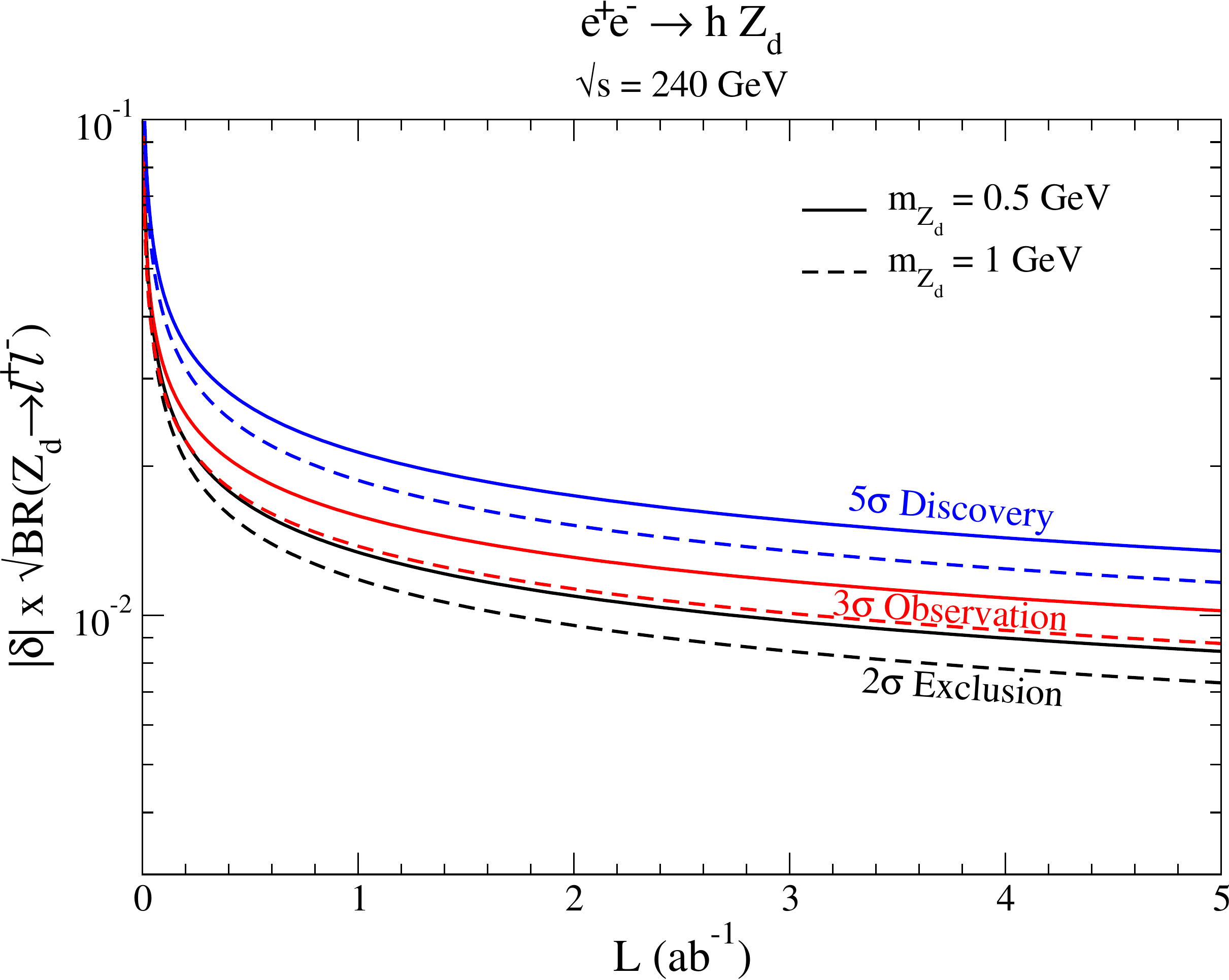}}
\subfigure[]{\includegraphics[width=0.45\textwidth,clip]{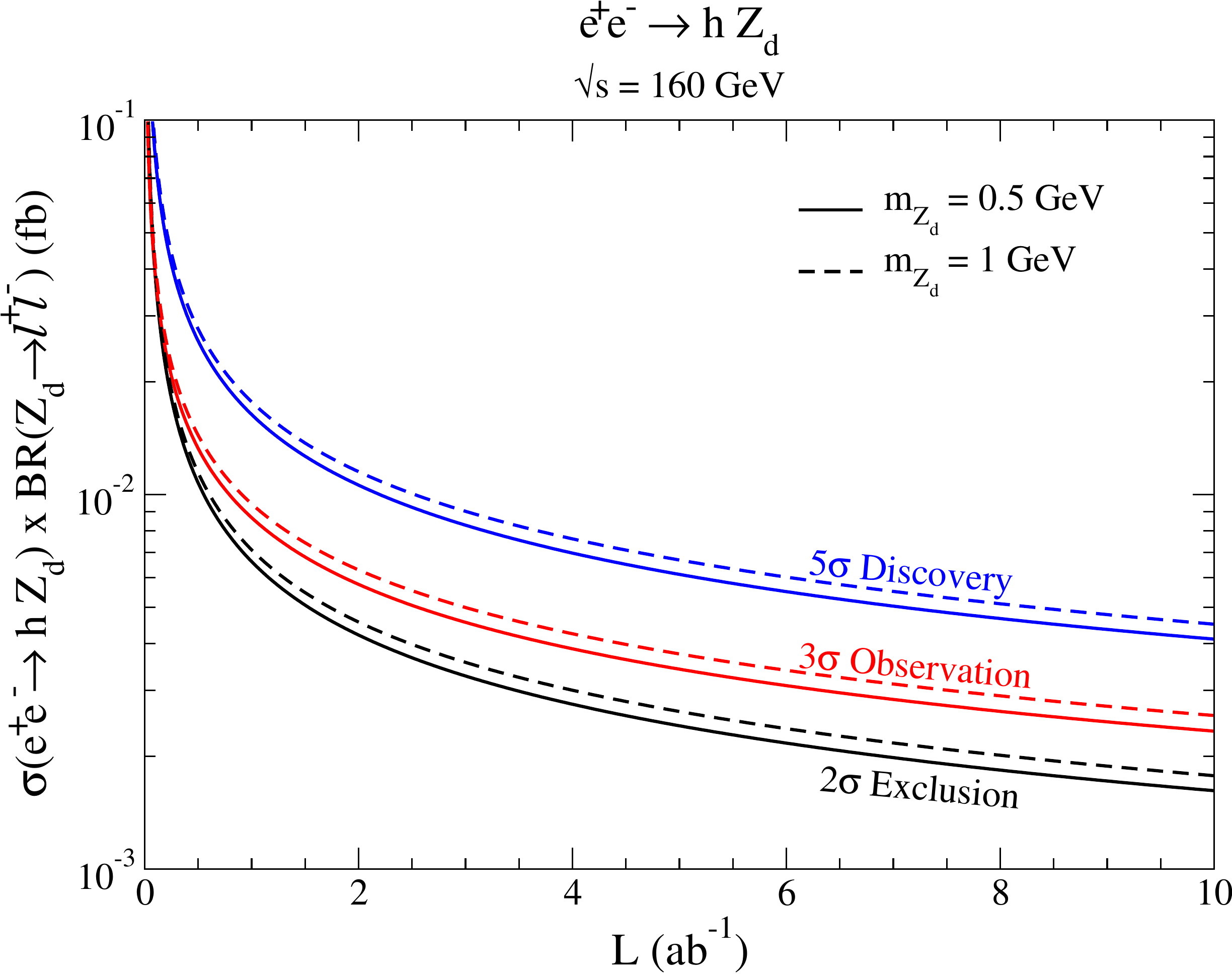}}
\subfigure[]{\includegraphics[width=0.45\textwidth,clip]{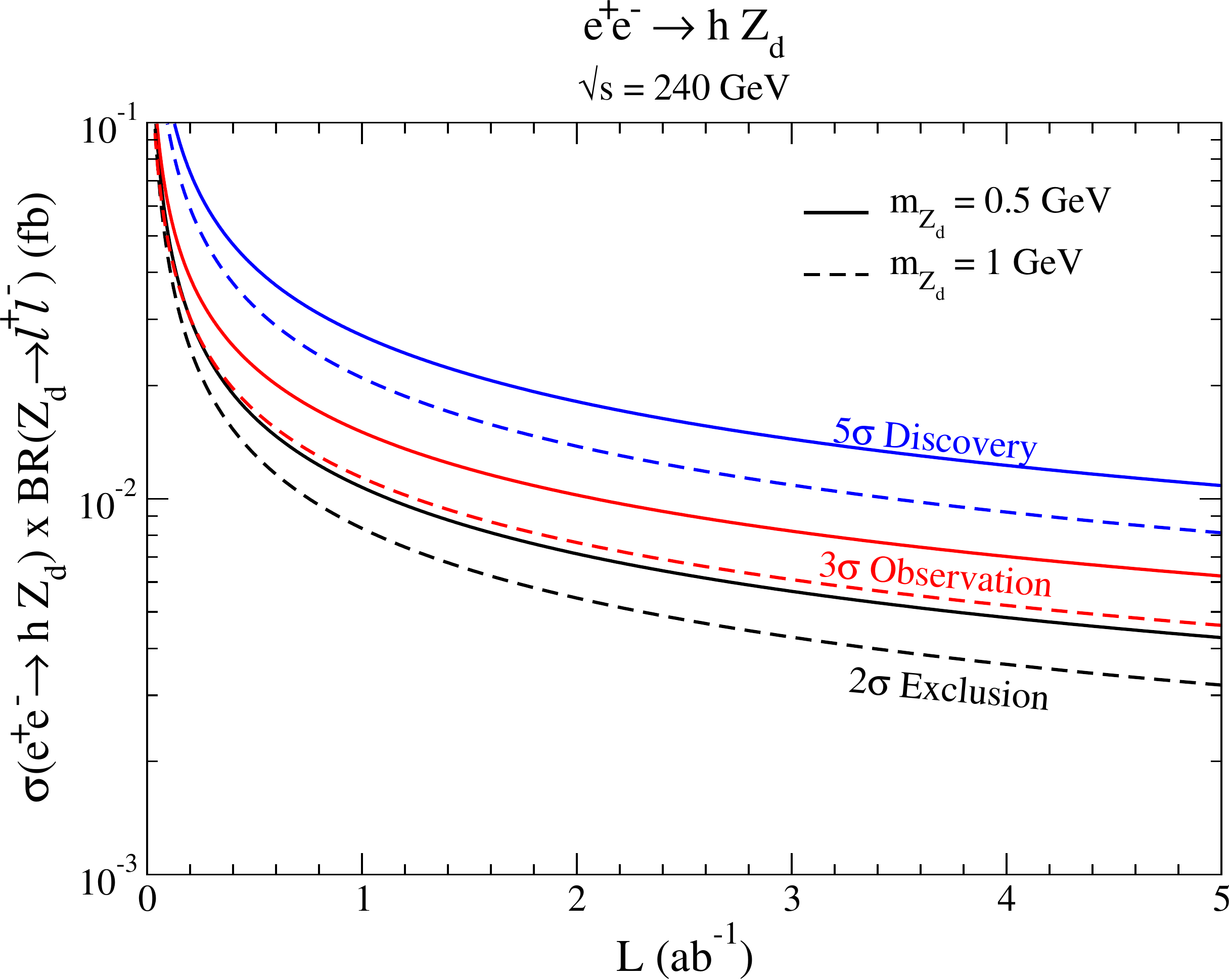}}
\end{center}
\caption{\label{fig:lumi} The luminosity needed for (black) exclusion [Eq.~(\ref{eq:exc})], (red) observation [Eq.~(\ref{eq:disc})], and (blue) discovery [Eq.~(\ref{eq:disc})] for either a given model parameter (a,b) $\delta\times \sqrt{{\rm BR}(Z_d\rightarrow \ell^+\ell^-)}$ or (c,d) cross section $\sigma(e^-e^+\rightarrow h\,Z_d){\rm BR}(Z_d\rightarrow \ell^+\ell^-)$.  These are shown at both (a,c) $\sqrt{s}=160$ GeV and (b,d) $\sqrt{s}=240$ GeV, and for $Z_d$ masses (solid) $m_{Z_d}=0.5$~GeV and (dashed) 1 GeV.  At least one $b$-tagged jet is required.  Here ${\rm BR}(Z_d\rightarrow \ell^+\ell^-)={\rm BR}(Z_d\rightarrow e^-e^+)+{\rm BR}(Z_d\rightarrow \mu^-\mu^+)$.}
\end{figure}

We also show the results if we require that at least one jet is tagged as a $b$-jet.  The $b$-tagging rate is set to 80\%, with a 1\% rate for light jets mis-tagged as $b$-jets and a 10\% mis-tagging rates for charm quarks~\cite{GuimaraesdaCosta:2678417}.  Requiring one $b$-jet increases the significance since the signal is mostly composed of $h\rightarrow b\bar{b}$ and the background has many processes with light and charm jets.  Since all signal and background processes are flavor conserving, requiring an additional $b$-tagged jet suppresses signal and background at similar rates.  We found that requiring two $b$-tagged jets does not help the significance.

In Fig.~\ref{fig:lumi} we show the estimated parameter regions that can be excluded at $2\sigma$, observed at $3\sigma$, and discovered at $5\sigma$.  Regions above the curves can be excluded, observed, or discovered. These are found requiring at least one $b$-tagged jet. Discovery and observation capability are defined using the discovery significances $\sigma_{\rm disc}\geq5$ and $\sigma_{\rm disc}\geq 3$, respectively.  To exclude, the signal plus background hypothesis is tested against the background only hypothesis~\cite{Cowan:2010js}:
\begin{eqnarray}
\sigma_{\rm exc}=\sqrt{-2\ln\left(\frac{L(S+B|B)}{L(B|B)}\right)}=\sqrt{2\left(S-B\log\left(1+\frac{S}{B}\right)\right)}.\label{eq:exc}
\end{eqnarray}
A signal is excluded at 2$\sigma$ when $\sigma_{\rm exc}\geq2$.    Note that Fig.~\ref{fig:Prod} shows that for a fixed $\delta$, the cross section of $e^+e^-\rightarrow h\,Z_d$ is lower at $\sqrt{s}=160$~GeV than at $\sqrt{s}=240$~GeV.  Hence, even though the parameter space reach is similar at both energies, the $\sqrt{s}=160$~GeV machine is sensitive to lower cross sections than $\sqrt{s}=240$~GeV.

\begin{table}[tb]
\newlength\wexp
\settowidth{\wexp}{$\sigma(e^-e^+\rightarrow h\,Z_d){\rm BR}(Z_d\rightarrow \ell^+\ell^-)$}
\newcolumntype{C}{>{\centering\arraybackslash}p{\dimexpr.333\wexp-\tabcolsep}}
\begin{center}
\resizebox{\textwidth}{!}{
\begin{tabular}{|c|c||c||ccc||CCC|}\hline
$\sqrt{s}$ & Luminosity &$m_{Z_d}$& \multicolumn{3}{c||}{$|\delta|\times\sqrt{{\rm BR}(Z_d\rightarrow \ell^+\ell^-)}$}& \multicolumn{3}{c|}{$\sigma(e^-e^+\rightarrow h\,Z_d){\rm BR}(Z_d\rightarrow \ell^+\ell^-)$}\\\cline{4-9}
 &  &  & 2$\sigma$ Exc. & 3$\sigma$ Obs. & 5$\sigma$ Disc. & 2$\sigma$ Exc. & 3$\sigma$ Obs. & 5$\sigma$ Disc.\\\hline
\multirow{2}{*}{160 GeV} & \multirow{2}{*}{10 ab$^{-1}$} & 0.5 GeV & $7.5\times 10^{-3}$ & $9.1\times 10^{-3}$& $1.2\times 10^{-2}$& 1.6 ab & 2.3 ab &4.1 ab \\
& & 1 GeV& $7.9\times 10^{-3}$ & $9.5\times 10^{-3}$ & $1.3\times10^{-2}$ & 1.8 ab & 2.6 ab & 4.5 ab\\\hline\hline
\multirow{2}{*}{240 GeV} & \multirow{2}{*}{5 ab$^{-1}$} & 0.5 GeV & $8.4\times 10^{-3}$ & $1.0\times 10^{-2}$& $1.3\times 10^{-2}$ & 4.3 ab & 6.2 ab &11 ab \\
& & 1 GeV& $7.3\times 10^{-3}$ & $8.8\times 10^{-3}$ & $1.2\times 10^{-2}$ & 3.2 ab & 4.6 ab & 8.1 ab\\\hline
\end{tabular}}
\caption{\label{tab:bounds} The smallest values of the parameter $\delta\times \sqrt{{\rm BR}(Z_d\rightarrow \ell^+\ell^-)}$ and cross section $\sigma(e^+e^-\rightarrow h\,Z_d){\rm BR}(Z_d\rightarrow\ell^+\ell^-$) that can be excluded at 2$\sigma$ (Exc.) [Eq.~(\ref{eq:exc})], observed at 3$\sigma$ (Obs.) [Eq.~(\ref{eq:disc})], or discovered at 5$\sigma$ (Disc.) [Eq.~(\ref{eq:disc})].  Results are shown for both masses $m_{Z_d}=0.5$~GeV and 1 GeV, and at both $\sqrt{s}=160$ and $240$ GeV with benchmark luminosities $10$ and $5$ ab$^{-1}$, respectively. At least one $b$-tagged jet is required and ${\rm BR}(Z_d\rightarrow \ell^+\ell^-)={\rm BR}(Z_d\rightarrow e^-e^+)+{\rm BR}(Z_d\rightarrow \mu^-\mu^+)$.}
\end{center}
\end{table}

The exclusion, observation, and discovery limits on model parameters and cross sections are given in Table~\ref{tab:bounds}.  With 10 ab$^{-1}$ at 160 GeV, an electron-positron machine could probe parameters down to $\delta\times\sqrt{{\rm BR}(Z_d\rightarrow \ell^+\ell^-)}\sim 8\times 10^{-3}$ and cross sections $\sigma(e^+e^-\rightarrow h\,Z_d){\rm BR}(Z_d\rightarrow \ell^+\ell^-)\sim 1-2$~ab.  The sensitivity of a 240 GeV machine with 5 ab$^{-1}$ is similar: $\delta\times\sqrt{{\rm BR}(Z_d\rightarrow \ell^+\ell^-)}\sim 8\times10^{-3}$ and cross sections $\sigma(e^+e^-\rightarrow h\,Z_d){\rm BR}(Z_d\rightarrow \ell^+\ell^-)\sim 3-4$~ab.

\section{Conclusions}
\label{sec:conc}
The particle physics community is going through discussions about future colliders beyond the LHC.  Many of these colliders have an extensive program to measure and search for new physics associated with the Higgs boson.  Due to the clean environment and large luminosities, electron-positron colliders will be able to measure the Higgs boson to great precision~\cite{CEPCStudyGroup:2018rmc,GuimaraesdaCosta:2678417,Behnke:2013xla,Baer:2013cma,Adolphsen:2013jya,Abada:2019zxq}.  Indeed, the European Strategy Group has endorsed an electron-positron collider as the next step in the global collider program~\cite{EuropeanStrategyGroup:2020pow}.

We propose a new search $e^+e^-\rightarrow h\,Z_d$, where $Z_d$ is a new light gauge boson.  As discussed in Secs.~\ref{sec:intro} and~\ref{sec:Model}, models with new light gauge bosons are theoretically well-motivated and there has been much effort in searching for these particles at high intensity, low energy experiments.  There have also been LHC searches for $h\rightarrow Z\,Z_d$, which arises from the same interactions as $e^+e^-\rightarrow h\,Z_d$.  However, as we discussed in Sec.~\ref{sec:Model} and~\ref{sec:collider}, the high resolutions of future detectors at electron-positron colliders will allow for searches with $Z_d$ masses below 1 GeV.  Current dedicated LHC searches do not probe masses this low~\cite{Aaboud:2018fvk,CMS-PAS-HIG-19-007}.

A particularly compelling argument for this search is that $e^+e^-\rightarrow h\,Z_d$ can be observed at collider energies below the $h\,Z$ threshold.  Indeed, many future electron-positron collider proposals include runs at the $WW$ threshold and our results indicate there could be discoverable new physics processes involving the Higgs boson at those energies.  Using the design luminosities of the FCC-ee, we showed that $e^+e^-\rightarrow h\,Z_d$ can be sensitive to parameters and cross sections of
\begin{center}
\begin{tabular}{lclcl}
$\delta\sim 8\times 10^{-3}$&{\rm and}& $\sigma(e^+e^-\rightarrow h\,Z_d)\sim 1-2$ ab~& at &$\sqrt{s}=160$ GeV with 10 ab$^{-1}$\\
$\delta\sim 8\times 10^{-3}$&{\rm and}& $\sigma(e^+e^-\rightarrow h\,Z_d)\sim 3-4$ ab~& at &$\sqrt{s}=240$ GeV with 5 ab$^{-1}$.
\end{tabular}
\end{center}

Once a discovery is made, it will be necessary to determine the origin of the $h-Z-Z_d$ coupling.  As discussed in Sec.~\ref{sec:Model}, by comparing the $e^+e^-\rightarrow h\,Z_d$ rate at different energies it will be possible to determine if this coupling originates from {dimension-5} or {dimension-3} operators.  Hence, even if $h\rightarrow Z\,Z_d$ is discovered first, searching for and measuring $e^+e^-\rightarrow h\,Z_d$ can provide important information about the new physics.  Additionally, if $h-Z-Z_d$ originates from the dimension-5 operator the $Z_d$ will be predominately transversely polarized, and if it originates from the dimension-3 operator the $Z_d$ will be predominately longitudinally polarized.  Hence, the angular distributions of the $Z_d$ decay products will also help disentangle the origin of these new interactions~\cite{Davoudiasl:2013aya}.

Finally, in our analysis we focused on the leptonic decays $Z_d\rightarrow \ell^+\ell^-$.  This choice was because the events, and therefore the signal, are fully reconstructable.  However, it is possible the the dark-$Z$ could decay into invisible particles, such as neutrinos or dark matter when kinematically allowed.  It would be interesting to further this study to include invisible $Z_d$ decays which would result in a Higgs and missing momentum signature even at 160~\rm{GeV}.  This could be promising for several reasons. (1) Even for hadronic decays the Higgs can be reconstructed well since the jet energy uncertainty is $\sim 4\%$ for Higgs decay products [Eq.~(\ref{eq:hadsmear})].  (2) At electron-positron colliders the complete missing energy and momentum vector are reconstructable.  Due to the two-body kinematics of our signal there is a precise prediction for the energy and the magnitude of that missing momentum vector [Eq.~(\ref{eq:Zdmom})].  Hence, it may still be possible efficiently tag signal events and search for invisible $Z_d$ decays.  It may even be feasible to search for higher $Z_d$ masses in ranges that are beyond the relevance of low energy constraints but still difficult for direct LHC searches due to the missing energy.

\section*{Acknowledgments}
IML would like to thank Hooman Davoudiasl for constructive comments on the manuscript and Hye-Sung Lee for encouraging discussions.  IML and YZ are supported in part by the U.S. Department of Energy under grant No. de-sc0019474.  PG is supported in part by the  State  of  Kansas  EPSCoR  grant  program.

\bibliographystyle{myutphys}
\bibliography{draft}

\providecommand{\href}[2]{#2}\begingroup\raggedright\begin{thebibliography}{10}

\bibitem{Arkani-Hamed:2015vfh}
N.~Arkani-Hamed, T.~Han, M.~Mangano, and L.-T. Wang, {\em {Physics
  opportunities of a 100 TeV proton\textendash{}proton collider}},
  \href{http://dx.doi.org/10.1016/j.physrep.2016.07.004}{{\em Phys. Rept.}
  {\bfseries 652} (2016) 1--49},
  \href{http://arxiv.org/abs/1511.06495}{{\ttfamily arXiv:1511.06495
  [hep-ph]}}.

\bibitem{Mangano:2016jyj}
M.~Mangano {\em et~al.}, {\em {Physics at a 100 TeV pp Collider: Standard Model
  Processes}}, \href{http://dx.doi.org/10.23731/CYRM-2017-003.1}{{\em CERN
  Yellow Rep.} no.~3, (2017) 1--254},
  \href{http://arxiv.org/abs/1607.01831}{{\ttfamily arXiv:1607.01831
  [hep-ph]}}.

\bibitem{Cepeda:2019klc}
M.~Cepeda {\em et~al.}, {\em {Report from Working Group 2}: {Higgs Physics at
  the HL-LHC and HE-LHC}},
  \href{http://dx.doi.org/10.23731/CYRM-2019-007.221}{{\em CERN Yellow Rep.
  Monogr.} {\bfseries 7} (2019) 221--584},
  \href{http://arxiv.org/abs/1902.00134}{{\ttfamily arXiv:1902.00134
  [hep-ph]}}.

\bibitem{deBlas:2019rxi}
J.~de~Blas {\em et~al.}, {\em {Higgs Boson Studies at Future Particle
  Colliders}}, \href{http://dx.doi.org/10.1007/JHEP01(2020)139}{{\em JHEP}
  {\bfseries 01} (2020) 139}, \href{http://arxiv.org/abs/1905.03764}{{\ttfamily
  arXiv:1905.03764 [hep-ph]}}.

\bibitem{Strategy:2019vxc}
R.~K. Ellis {\em et~al.}, {\em {Physics Briefing Book}: {Input for the European
  Strategy for Particle Physics Update 2020}},
  \href{http://arxiv.org/abs/1910.11775}{{\ttfamily arXiv:1910.11775
  [hep-ex]}}.

\bibitem{Bechtle:2014ewa}
P.~Bechtle, S.~Heinemeyer, O.~St\r{a}l, T.~Stefaniak, and G.~Weiglein, {\em
  {Probing the Standard Model with Higgs signal rates from the Tevatron, the
  LHC and a future ILC}}, \href{http://dx.doi.org/10.1007/JHEP11(2014)039}{{\em
  JHEP} {\bfseries 11} (2014) 039},
  \href{http://arxiv.org/abs/1403.1582}{{\ttfamily arXiv:1403.1582 [hep-ph]}}.

\bibitem{Liu:2016zki}
Z.~Liu, L.-T. Wang, and H.~Zhang, {\em {Exotic decays of the 125 GeV Higgs
  boson at future $e^+e^-$ lepton colliders}},
  \href{http://dx.doi.org/10.1088/1674-1137/41/6/063102}{{\em Chin. Phys. C}
  {\bfseries 41} no.~6, (2017) 063102},
  \href{http://arxiv.org/abs/1612.09284}{{\ttfamily arXiv:1612.09284
  [hep-ph]}}.

\bibitem{Gu:2017ckc}
J.~Gu, H.~Li, Z.~Liu, S.~Su, and W.~Su, {\em {Learning from Higgs Physics at
  Future Higgs Factories}},
  \href{http://dx.doi.org/10.1007/JHEP12(2017)153}{{\em JHEP} {\bfseries 12}
  (2017) 153}, \href{http://arxiv.org/abs/1709.06103}{{\ttfamily
  arXiv:1709.06103 [hep-ph]}}.

\bibitem{Durieux:2017rsg}
G.~Durieux, C.~Grojean, J.~Gu, and K.~Wang, {\em {The leptonic future of the
  Higgs}}, \href{http://dx.doi.org/10.1007/JHEP09(2017)014}{{\em JHEP}
  {\bfseries 09} (2017) 014}, \href{http://arxiv.org/abs/1704.02333}{{\ttfamily
  arXiv:1704.02333 [hep-ph]}}.

\bibitem{Barklow:2017suo}
T.~Barklow, K.~Fujii, S.~Jung, R.~Karl, J.~List, T.~Ogawa, M.~E. Peskin, and
  J.~Tian, {\em {Improved Formalism for Precision Higgs Coupling Fits}},
  \href{http://dx.doi.org/10.1103/PhysRevD.97.053003}{{\em Phys. Rev. D}
  {\bfseries 97} no.~5, (2018) 053003},
  \href{http://arxiv.org/abs/1708.08912}{{\ttfamily arXiv:1708.08912
  [hep-ph]}}.

\bibitem{Voigt:2017vfz}
A.~Voigt and S.~Westhoff, {\em {Virtual signatures of dark sectors in Higgs
  couplings}}, \href{http://dx.doi.org/10.1007/JHEP11(2017)009}{{\em JHEP}
  {\bfseries 11} (2017) 009}, \href{http://arxiv.org/abs/1708.01614}{{\ttfamily
  arXiv:1708.01614 [hep-ph]}}.

\bibitem{An:2018dwb}
F.~An {\em et~al.}, {\em {Precision Higgs physics at the CEPC}},
  \href{http://dx.doi.org/10.1088/1674-1137/43/4/043002}{{\em Chin. Phys. C}
  {\bfseries 43} no.~4, (2019) 043002},
  \href{http://arxiv.org/abs/1810.09037}{{\ttfamily arXiv:1810.09037
  [hep-ex]}}.

\bibitem{deBlas:2018mhx}
R.~Franceschini {\em et~al.}, {\em {The CLIC Potential for New Physics}},
  \href{http://arxiv.org/abs/1812.02093}{{\ttfamily arXiv:1812.02093
  [hep-ph]}}.

\bibitem{deBlas:2019wgy}
J.~De~Blas, G.~Durieux, C.~Grojean, J.~Gu, and A.~Paul, {\em {On the future of
  Higgs, electroweak and diboson measurements at lepton colliders}},
  \href{http://dx.doi.org/10.1007/JHEP12(2019)117}{{\em JHEP} {\bfseries 12}
  (2019) 117}, \href{http://arxiv.org/abs/1907.04311}{{\ttfamily
  arXiv:1907.04311 [hep-ph]}}.

\bibitem{Tan:2020fxk}
Y.~Tan, X.~Shi, R.~Kiuchi, M.~Ruan, M.~Jing, X.~Mo, X.~Lou, G.~Li, K.~Zhang,
  and S.~Jyotishmati, {\em {Search for invisible decay of a Higgs boson
  produced at the CEPC}}, \href{http://arxiv.org/abs/2001.05912}{{\ttfamily
  arXiv:2001.05912 [hep-ex]}}.

\bibitem{Jung:2020uzh}
S.~Jung, J.~Lee, M.~Perell\'o, J.~Tian, and M.~Vos, {\em {Higgs, top and
  electro-weak precision measurements at future $e^+ e^-$ colliders; a combined
  effective field theory analysis with renormalization mixing}},
  \href{http://arxiv.org/abs/2006.14631}{{\ttfamily arXiv:2006.14631
  [hep-ph]}}.

\bibitem{Fuchs:2020cmm}
E.~Fuchs, O.~Matsedonskyi, I.~Savoray, and M.~Schlaffer, {\em {Collider
  searches of scalar singlets across lifetimes}},
  \href{http://arxiv.org/abs/2008.12773}{{\ttfamily arXiv:2008.12773
  [hep-ph]}}.

\bibitem{Li:2020glc}
H.~Li, H.~Song, S.~Su, W.~Su, and J.~M. Yang, {\em {MSSM at future Higgs
  factories}}, \href{http://arxiv.org/abs/2010.09782}{{\ttfamily
  arXiv:2010.09782 [hep-ph]}}.

\bibitem{Drechsel:2018mgd}
P.~Drechsel, G.~Moortgat-Pick, and G.~Weiglein, {\em {Prospects for direct
  searches for light Higgs bosons at the ILC with 250 GeV}},
  \href{http://dx.doi.org/10.1140/epjc/s10052-020-08438-1}{{\em Eur. Phys. J.
  C} {\bfseries 80} no.~10, (2020) 922},
  \href{http://arxiv.org/abs/1801.09662}{{\ttfamily arXiv:1801.09662
  [hep-ph]}}.

\bibitem{Kalinowski:2018kdn}
J.~Kalinowski, W.~Kotlarski, T.~Robens, D.~Sokolowska, and A.~F. Zarnecki, {\em
  {Exploring Inert Scalars at CLIC}},
  \href{http://dx.doi.org/10.1007/JHEP07(2019)053}{{\em JHEP} {\bfseries 07}
  (2019) 053}, \href{http://arxiv.org/abs/1811.06952}{{\ttfamily
  arXiv:1811.06952 [hep-ph]}}.

\bibitem{Bahl:2020kwe}
H.~Bahl, P.~Bechtle, S.~Heinemeyer, S.~Liebler, T.~Stefaniak, and G.~Weiglein,
  {\em {HL-LHC and ILC sensitivities in the hunt for heavy Higgs bosons}},
  \href{http://dx.doi.org/10.1140/epjc/s10052-020-08472-z}{{\em Eur. Phys. J.
  C} {\bfseries 80} no.~10, (2020) 916},
  \href{http://arxiv.org/abs/2005.14536}{{\ttfamily arXiv:2005.14536
  [hep-ph]}}.

\bibitem{CEPCStudyGroup:2018rmc}
{\bfseries CEPC Study Group} Collaboration, {\em {CEPC Conceptual Design
  Report: Volume 1 - Accelerator}},
  \href{http://arxiv.org/abs/1809.00285}{{\ttfamily arXiv:1809.00285
  [physics.acc-ph]}}.

\bibitem{GuimaraesdaCosta:2678417}
J.~a.~B. Guimarães~da Costa {\em et~al.}, , {\bfseries CEPC Study Group
  Collaboration} Collaboration, {\em {CEPC Conceptual Design Report: Volume 2 -
  Physics \& Detector}}, tech. rep., Nov, 2018.
\newblock \href{http://arxiv.org/abs/1811.10545}{{\ttfamily arXiv:1811.10545
  [hep-ex]}}.
\newblock 424 pages.

\bibitem{Behnke:2013xla}
{\em {The International Linear Collider Technical Design Report - Volume 1:
  Executive Summary}}, \href{http://arxiv.org/abs/1306.6327}{{\ttfamily
  arXiv:1306.6327 [physics.acc-ph]}}.

\bibitem{Baer:2013cma}
{\em {The International Linear Collider Technical Design Report - Volume 2:
  Physics}}, \href{http://arxiv.org/abs/1306.6352}{{\ttfamily arXiv:1306.6352
  [hep-ph]}}.

\bibitem{Adolphsen:2013jya}
{\em {The International Linear Collider Technical Design Report - Volume 3.I:
  Accelerator \textbackslash{}\& in the Technical Design Phase}},
  \href{http://arxiv.org/abs/1306.6353}{{\ttfamily arXiv:1306.6353
  [physics.acc-ph]}}.

\bibitem{Abada:2019zxq}
A.~Abada {\em et~al.}, , {\bfseries FCC} Collaboration, {\em {FCC-ee: The
  Lepton Collider}: {Future Circular Collider Conceptual Design Report Volume
  2}}, \href{http://dx.doi.org/10.1140/epjst/e2019-900045-4}{{\em Eur. Phys. J.
  ST} {\bfseries 228} no.~2, (2019) 261--623}.

\bibitem{Langacker:2008yv}
P.~Langacker, {\em {The Physics of Heavy $Z^\prime$ Gauge Bosons}},
  \href{http://dx.doi.org/10.1103/RevModPhys.81.1199}{{\em Rev. Mod. Phys.}
  {\bfseries 81} (2009) 1199--1228},
  \href{http://arxiv.org/abs/0801.1345}{{\ttfamily arXiv:0801.1345 [hep-ph]}}.

\bibitem{Jaeckel:2010ni}
J.~Jaeckel and A.~Ringwald, {\em {The Low-Energy Frontier of Particle
  Physics}}, \href{http://dx.doi.org/10.1146/annurev.nucl.012809.104433}{{\em
  Ann. Rev. Nucl. Part. Sci.} {\bfseries 60} (2010) 405--437},
  \href{http://arxiv.org/abs/1002.0329}{{\ttfamily arXiv:1002.0329 [hep-ph]}}.

\bibitem{Hewett:2012ns}
\href{http://dx.doi.org/10.2172/1042577}{{\em {Fundamental Physics at the
  Intensity Frontier}}}.
\newblock 5, 2012.
\newblock \href{http://arxiv.org/abs/1205.2671}{{\ttfamily arXiv:1205.2671
  [hep-ex]}}.

\bibitem{Essig:2013lka}
R.~Essig {\em et~al.}, {\em {Working Group Report: New Light Weakly Coupled
  Particles}}, in {\em {Community Summer Study 2013}: {Snowmass on the
  Mississippi}}.
\newblock 10, 2013.
\newblock \href{http://arxiv.org/abs/1311.0029}{{\ttfamily arXiv:1311.0029
  [hep-ph]}}.

\bibitem{Alexander:2016aln}
J.~Alexander {\em et~al.}, {\em {Dark Sectors 2016 Workshop: Community
  Report}},
\newblock 8, 2016.
\newblock \href{http://arxiv.org/abs/1608.08632}{{\ttfamily arXiv:1608.08632
  [hep-ph]}}.

\bibitem{Battaglieri:2017aum}
M.~Battaglieri {\em et~al.}, {\em {US Cosmic Visions: New Ideas in Dark Matter
  2017: Community Report}}, in {\em {U.S. Cosmic Visions: New Ideas in Dark
  Matter}}.
\newblock 7, 2017.
\newblock \href{http://arxiv.org/abs/1707.04591}{{\ttfamily arXiv:1707.04591
  [hep-ph]}}.

\bibitem{Holdom:1985ag}
B.~Holdom, {\em {Two U(1)'s and Epsilon Charge Shifts}},
  \href{http://dx.doi.org/10.1016/0370-2693(86)91377-8}{{\em Phys. Lett. B}
  {\bfseries 166} (1986) 196--198}.

\bibitem{Galison:1983pa}
P.~Galison and A.~Manohar, {\em {TWO Z's OR NOT TWO Z's?}},
  \href{http://dx.doi.org/10.1016/0370-2693(84)91161-4}{{\em Phys. Lett. B}
  {\bfseries 136} (1984) 279--283}.

\bibitem{Dienes:1996zr}
K.~R. Dienes, C.~F. Kolda, and J.~March-Russell, {\em {Kinetic mixing and the
  supersymmetric gauge hierarchy}},
  \href{http://dx.doi.org/10.1016/S0550-3213(97)00173-9}{{\em Nucl. Phys. B}
  {\bfseries 492} (1997) 104--118},
  \href{http://arxiv.org/abs/hep-ph/9610479}{{\ttfamily arXiv:hep-ph/9610479}}.

\bibitem{Bjorken:2009mm}
J.~D. Bjorken, R.~Essig, P.~Schuster, and N.~Toro, {\em {New Fixed-Target
  Experiments to Search for Dark Gauge Forces}},
  \href{http://dx.doi.org/10.1103/PhysRevD.80.075018}{{\em Phys. Rev. D}
  {\bfseries 80} (2009) 075018},
  \href{http://arxiv.org/abs/0906.0580}{{\ttfamily arXiv:0906.0580 [hep-ph]}}.

\bibitem{Davoudiasl:2012qa}
H.~Davoudiasl, H.-S. Lee, and W.~J. Marciano, {\em {Muon Anomaly and Dark
  Parity Violation}},
  \href{http://dx.doi.org/10.1103/PhysRevLett.109.031802}{{\em Phys. Rev.
  Lett.} {\bfseries 109} (2012) 031802},
  \href{http://arxiv.org/abs/1205.2709}{{\ttfamily arXiv:1205.2709 [hep-ph]}}.

\bibitem{Davoudiasl:2013aya}
H.~Davoudiasl, H.-S. Lee, I.~Lewis, and W.~J. Marciano, {\em {Higgs Decays as a
  Window into the Dark Sector}},
  \href{http://dx.doi.org/10.1103/PhysRevD.88.015022}{{\em Phys. Rev. D}
  {\bfseries 88} no.~1, (2013) 015022},
  \href{http://arxiv.org/abs/1304.4935}{{\ttfamily arXiv:1304.4935 [hep-ph]}}.

\bibitem{Davoudiasl:2012ag}
H.~Davoudiasl, H.-S. Lee, and W.~J. Marciano, {\em {'Dark' Z implications for
  Parity Violation, Rare Meson Decays, and Higgs Physics}},
  \href{http://dx.doi.org/10.1103/PhysRevD.85.115019}{{\em Phys. Rev. D}
  {\bfseries 85} (2012) 115019},
  \href{http://arxiv.org/abs/1203.2947}{{\ttfamily arXiv:1203.2947 [hep-ph]}}.

\bibitem{Davoudiasl:2015bua}
H.~Davoudiasl, H.-S. Lee, and W.~J. Marciano, {\em {Low $Q^2$ weak mixing angle
  measurements and rare Higgs decays}},
  \href{http://dx.doi.org/10.1103/PhysRevD.92.055005}{{\em Phys. Rev. D}
  {\bfseries 92} no.~5, (2015) 055005},
  \href{http://arxiv.org/abs/1507.00352}{{\ttfamily arXiv:1507.00352
  [hep-ph]}}.

\bibitem{Curtin:2013fra}
D.~Curtin {\em et~al.}, {\em {Exotic decays of the 125 GeV Higgs boson}},
  \href{http://dx.doi.org/10.1103/PhysRevD.90.075004}{{\em Phys. Rev. D}
  {\bfseries 90} no.~7, (2014) 075004},
  \href{http://arxiv.org/abs/1312.4992}{{\ttfamily arXiv:1312.4992 [hep-ph]}}.

\bibitem{Curtin:2014cca}
D.~Curtin, R.~Essig, S.~Gori, and J.~Shelton, {\em {Illuminating Dark Photons
  with High-Energy Colliders}},
  \href{http://dx.doi.org/10.1007/JHEP02(2015)157}{{\em JHEP} {\bfseries 02}
  (2015) 157}, \href{http://arxiv.org/abs/1412.0018}{{\ttfamily arXiv:1412.0018
  [hep-ph]}}.

\bibitem{Gopalakrishna:2008dv}
S.~Gopalakrishna, S.~Jung, and J.~D. Wells, {\em {Higgs boson decays to four
  fermions through an abelian hidden sector}},
  \href{http://dx.doi.org/10.1103/PhysRevD.78.055002}{{\em Phys. Rev. D}
  {\bfseries 78} (2008) 055002},
  \href{http://arxiv.org/abs/0801.3456}{{\ttfamily arXiv:0801.3456 [hep-ph]}}.

\bibitem{Aaboud:2018fvk}
M.~Aaboud {\em et~al.}, , {\bfseries ATLAS} Collaboration, {\em {Search for
  Higgs boson decays to beyond-the-Standard-Model light bosons in four-lepton
  events with the ATLAS detector at $\sqrt{s}=13$ TeV}},
  \href{http://dx.doi.org/10.1007/JHEP06(2018)166}{{\em JHEP} {\bfseries 06}
  (2018) 166}, \href{http://arxiv.org/abs/1802.03388}{{\ttfamily
  arXiv:1802.03388 [hep-ex]}}.

\bibitem{CMS-PAS-HIG-19-007}
{\bfseries CMS Collaboration} Collaboration, {\em {Search for a low-mass
  dilepton resonance in Higgs boson decays to four-lepton final states at
  $\sqrt{s}=13~\mathrm{TeV}$}}, Tech. Rep. CMS-PAS-HIG-19-007, CERN, Geneva,
  2020.

\bibitem{Davoudiasl:2012ig}
H.~Davoudiasl, H.-S. Lee, and W.~J. Marciano, {\em {Dark Side of Higgs Diphoton
  Decays and Muon g-2}},
  \href{http://dx.doi.org/10.1103/PhysRevD.86.095009}{{\em Phys. Rev. D}
  {\bfseries 86} (2012) 095009},
  \href{http://arxiv.org/abs/1208.2973}{{\ttfamily arXiv:1208.2973 [hep-ph]}}.

\bibitem{Davoudiasl:2014kua}
H.~Davoudiasl, H.-S. Lee, and W.~J. Marciano, {\em {Muon $g−2$, rare kaon
  decays, and parity violation from dark bosons}},
  \href{http://dx.doi.org/10.1103/PhysRevD.89.095006}{{\em Phys. Rev. D}
  {\bfseries 89} no.~9, (2014) 095006},
  \href{http://arxiv.org/abs/1402.3620}{{\ttfamily arXiv:1402.3620 [hep-ph]}}.

\bibitem{Heinemeyer:2013tqa}
J.~R. Andersen {\em et~al.}, , {\bfseries LHC Higgs Cross Section Working
  Group} Collaboration, {\em {Handbook of LHC Higgs Cross Sections: 3. Higgs
  Properties}}, \href{http://arxiv.org/abs/1307.1347}{{\ttfamily
  arXiv:1307.1347 [hep-ph]}}.

\bibitem{deFlorian:2016spz}
D.~de~Florian {\em et~al.}, , {\bfseries LHC Higgs Cross Section Working Group}
  Collaboration, {\em {Handbook of LHC Higgs Cross Sections: 4. Deciphering the
  Nature of the Higgs Sector}},
  \href{http://arxiv.org/abs/1610.07922}{{\ttfamily arXiv:1610.07922
  [hep-ph]}}.

\bibitem{Alwall:2014hca}
J.~Alwall, R.~Frederix, S.~Frixione, V.~Hirschi, F.~Maltoni, O.~Mattelaer,
  H.~S. Shao, T.~Stelzer, P.~Torrielli, and M.~Zaro, {\em {The automated
  computation of tree-level and next-to-leading order differential cross
  sections, and their matching to parton shower simulations}},
  \href{http://dx.doi.org/10.1007/JHEP07(2014)079}{{\em JHEP} {\bfseries 07}
  (2014) 079}, \href{http://arxiv.org/abs/1405.0301}{{\ttfamily arXiv:1405.0301
  [hep-ph]}}.

\bibitem{Christensen:2008py}
N.~D. Christensen and C.~Duhr, {\em {FeynRules - Feynman rules made easy}},
  \href{http://dx.doi.org/10.1016/j.cpc.2009.02.018}{{\em Comput. Phys.
  Commun.} {\bfseries 180} (2009) 1614--1641},
  \href{http://arxiv.org/abs/0806.4194}{{\ttfamily arXiv:0806.4194 [hep-ph]}}.

\bibitem{Alloul:2013bka}
A.~Alloul, N.~D. Christensen, C.~Degrande, C.~Duhr, and B.~Fuks, {\em
  {FeynRules 2.0 - A complete toolbox for tree-level phenomenology}},
  \href{http://dx.doi.org/10.1016/j.cpc.2014.04.012}{{\em Comput. Phys.
  Commun.} {\bfseries 185} (2014) 2250--2300},
  \href{http://arxiv.org/abs/1310.1921}{{\ttfamily arXiv:1310.1921 [hep-ph]}}.

\bibitem{Cowan:2010js}
G.~Cowan, K.~Cranmer, E.~Gross, and O.~Vitells, {\em {Asymptotic formulae for
  likelihood-based tests of new physics}},
  \href{http://dx.doi.org/10.1140/epjc/s10052-011-1554-0}{{\em Eur. Phys. J. C}
  {\bfseries 71} (2011) 1554}, \href{http://arxiv.org/abs/1007.1727}{{\ttfamily
  arXiv:1007.1727 [physics.data-an]}}. [Erratum: Eur.Phys.J.C 73, 2501 (2013)].

\bibitem{EuropeanStrategyGroup:2020pow}
{\bfseries European Strategy~Group} Collaboration, ,
  \href{http://dx.doi.org/10.17181/ESU2020}{{\em {2020 Update of the European
  Strategy for Particle Physics}}}.
\newblock CERN Council, Geneva, 2020.

\end{thebibliography}\endgroup

\end{document}